\documentclass[namedreferences,hyperref,optionalrh]{spr-sola}%
\usepackage{color}           
\usepackage{lineno,xcolor,multirow,array,booktabs,amssymb,amsmath,natbib}%
\usepackage{graphicx,epstopdf,hyperref,appendix,threeparttable,chngpage}
\usepackage{txfonts}

\renewcommand{\vec}[1]{{\mathbfit #1}}

\newcommand{\rsun}{R$_{\odot}$}
\newcommand{\kms}{km~s$^{-1}$}
\newcommand{\hi}{H \textsc{i} }
\newcommand{\heii}{He \textsc{ii} }
\newcommand{\Sixi}{Si \textsc{xi} }
\newcommand{\lya}{Ly$\alpha$ }
\newcommand{\alfven}{Alfv\'{e}n}

\chardef\us=`\_

\begin{document}

\begin{frontmatter}
\title{Parameter effects on the total intensity of \hi \lya line for a modelled coronal mass ejection and its driven shock}

\author[addressref={aff1},corref,email={yingbl@pmo.ac.cn}]{\inits{B.L.}\fnm{Beili}~\snm{Ying}\orcid{0000-0001-8402-9748}}
\author[addressref={aff1,aff2},email={shigl@pmo.ac.cn}]{\inits{G.L.}\fnm{Guanglu}~\snm{Shi}\orcid{0000-0001-7397-455X}}
\author[addressref={aff1,aff2},corref,email={lfeng@pmo.ac.cn}]{\inits{L.}\fnm{Li}~\snm{Feng}\orcid{0000-0003-4655-6939}}
\author[addressref=aff1,email={leilu@pmo.ac.cn}]{\inits{L.}\fnm{Lei}~\snm{Lu}\orcid{0000-0002-3032-6066}}
\author[addressref=aff1,email={xuejc@pmo.ac.cn}]{\inits{J.C.}\fnm{Jianchao}~\snm{Xue}\orcid{0000-0003-4829-9067}}
\author[addressref={aff1,aff2},email={lisht@pmo.ac.cn}]{\inits{S.T.}\fnm{Shuting}~\snm{Li}\orcid{0000-0003-2694-2875}}
\author[addressref={aff1,aff3},email={wqgan@pmo.ac.cn}]{\inits{W.Q.}\fnm{Weiqun}~\snm{Gan}}
\author[addressref={aff1,aff2},email={nj.lihui@pmo.ac.cn}]{\inits{H.}\fnm{Hui}~\snm{Li}\orcid{0000-0003-1078-3021}}

\address[id=aff1]{Key Laboratory of Dark Matter and Space Astronomy, Purple Mountain Observatory, Chinese Academy of Sciences, Nanjing 210023, People's Republic of China}
\address[id=aff2]{School of Astronomy and Space Science, University of Science and Technology of China, Hefei, Anhui 230026, People's Republic of China}
\address[id=aff3]{University of Chinese Academy of Sciences, Nanjing 211135, People's Republic of China}

\runningauthor{Ying et al.}
\runningtitle{Parameter effects on the total intensity of \hi \lya line for a modeled CME and its driven shock}

\begin{abstract}
The combination of the \hi \lya (121.6 nm) line formation mechanism with ultraviolet (UV) \lya and white-light (WL) observations provides an effective method for determining the electron temperature of coronal mass ejections (CMEs). A key to ensuring the accuracy of this diagnostic technique is the precise calculation of theoretical \lya intensities. This study performs a modelled CME and its driven shock via the three-dimensional numerical magneto-hydrodynamic simulation. Then, we generate synthetic UV and WL images of the CME and shock within a few solar radii to quantify the impact of different assumptions on the theoretical \lya intensities, such as the incident intensity of the solar chromospheric \lya line ($I_{disk}$), the geometric scattering function ($p(\theta)$), and the kinetic temperature ($T_{\boldsymbol{n}}$) assumed to be equal to either the proton ($T_p$) or electron ($T_e$) temperatures. By comparing differences of the \lya intensities of the CME and shock under these assumptions, we find that: (1) Using the uniform or Carrington maps of the disk \lya emission underestimates the corona \lya intensity (with relative uncertainties below 10\%) compared to the synchronic map, except for a slight overestimate ($<4$\%) observed in the partial CME core. The Carrington map yields lower uncertainties than the uniform disk. (2) Neglecting the geometric scattering process has a relatively minor impact on the \lya intensity, with a maximum relative uncertainty of no more than 5\%. The \lya intensity is underestimated for the most part but overestimated in the CME core. (3) Compared to the assumption $T_{\boldsymbol{n}}=T_p$, using $T_{\boldsymbol{n}}=T_e$ leads to more complex relative uncertainties in CME \lya intensity. The CME core and void are both overestimated, with the maximum relative uncertainty in the core exceeding 50\% and in the void remaining below 35\%. An appropriate increasing proton-to-electron temperature ratio can reduce the uncertainty in the CME core and void. In the CME front, both overestimates and underestimates exist with relative uncertainties of less than 35\%. The electron temperature assumption has a smaller impact on the shock, with an underestimated relative uncertainty of less than 20\%.
\end{abstract}
\end{frontmatter}
\section{Introduction}
     \label{S-Introduction} 

Coronal mass ejections (CMEs) are highly energetic eruptions that occur in the solar atmosphere. Various instruments have been used to observe CMEs, revealing their distinct characteristics in different wavelengths. When the velocity of a CME surpasses the local fast magnetosonic velocity, it can drive a shock wave \citep[e.g.][]{Gopalswamy2012, Ying2018}.

Several studies have demonstrated the valuable insights gained by combining white-light (WL) coronagraphic images with ultraviolet (UV) observations from the UV Coronagraph Spectrometer \citep[UVCS,][]{Kohl1995}. This combination offers a unique capability to investigate various aspects of CMEs, including plasma temperatures, kinematic properties, and elemental distributions \citep{Ciaravella2003, Ciaravella2006, Raymond2003, Bemporad2007, Ying2020,Bemporadsym2022}. Furthermore, it enables the study of numerous CME-related phenomena, such as CME-driven shocks \citep{Ciaravella2005, Ciaravella2006, Bemporad2010}, post-CME current sheets \citep{Bemporad2008, Shigl2020, Bemporad2024}, and CME-driven reconnections \citep{Bemporad2010}. In UV \lya images, fast CMEs can exhibit a dark front caused by the Doppler dimming (DD) effect \citep{Bemporad2018, Ying2020}. This effect arises from the Doppler shift of the coronal absorption profile relative to the solar disk incident profile, resulting in a reduced efficiency of atomic excitation according to the formation mechanism of the \hi \lya line. The DD technique has been developed to determine solar wind speed \citep{Noci1987, Dolei2018, Dolei2019}, and it has also been applied to estimate the electron temperature of CME plasma, even without slit-spectroscopic data, by combining WL and UV \hi \lya line intensities \citep{Susino2016, Bemporad2018, Ying2020,Bemporadsym2022}. The crucial aspect of this method lies in acquiring an accurate calculation of the theoretical \lya intensity. By comparing the observed \lya intensity with the theoretical value, the electron temperature of the CME can be determined. However, many of these studies simplify the calculation of the theoretical \lya intensity in the DD method by (1) assuming a uniform incident \lya intensity spectrum from the solar disk, (2) neglecting the geometric scattering process and treating the Sun as a point source, and (3) considering the neutral hydrogen kinetic temperature to be equal to the electron temperature in the absence of spectroscopic observations. These simplifications can potentially lead to inaccuracies in the theoretical \lya intensity, consequently impacting the accuracy of the estimated electron temperatures. 

Several new instruments have been developed to enable simultaneous imaging of the solar corona using WL and UV \hi \lya lines, eliminating the need for slit-spectroscopic observations. Notable examples include the \lya Solar Telescope \citep[LST,][]{Li2019, Feng2019} on board the Chinese Advanced Space-based Solar Observatory mission \citep[ASO-S,][]{Gan2019,Gan2023}, and the Metis coronagraph on board the ESA Solar Orbiter mission \citep{Antonucci2017, Antonucci2020}. The LST comprises three instruments: a Solar Disk Imager (SDI) capable of imaging the solar disk up to 1.2 \rsun~in the \lya line, a White-light Solar Telescope (WST), and a Solar Corona Imager (SCI) with a field-of-view (FoV) ranging from 1.1 to 2.5 \rsun~in both WL and \lya passbands. The Metis coronagraph has a FoV of $1.6^{\circ}–2.9^{\circ}$ (corresponding to projected altitudes from 1.7 to 3.1 \rsun when the spacecraft reaches its closest approach at 0.28 AU). With the deployment of these new instruments, the DD method becomes a valuable tool for diagnosing the physical parameters of CMEs and shocks. However, when the instrument's FoV approaches the Sun, particularly in the case of LST, it becomes necessary to investigate the effects of geometric scattering angles and incident chromospheric \hi \lya intensity on the total \lya radiation from CMEs and shocks when applying the DD technique. We can consider that the most \lya intensity originates from the chromosphere, due to the narrow width of the reversal center of the \lya line formed around the transition region, and the fact that the line wings are formed in the mid-chromosphere \citep{Vernazza1981, Hong2019ApJ}. In a study by \citet{Auchere2005a}, they investigated the impact of chromospheric \lya intensity anisotropy on the total \lya intensity of the corona due to resonant scattering. They discovered that during the solar minimum, assuming a uniform disk distribution leads to a systematic overestimation of the total intensity in the polar regions by an average of 15\%. Additionally, \citet{Dolei2019} examined the influence of non-uniform chromospheric \lya radiation on determining the \hi outflow velocity in the corona. They found that assuming a uniform chromospheric intensity underestimates the irradiance in the equatorial regions, with a maximum underestimation of 15\% at 1.5 \rsun, decreasing to 5\% at 4 \rsun. Conversely, in the solar polar regions, the total irradiance is overestimated. However, the impact of these assumptions on the \hi \lya intensities of CMEs and their driven shocks still remains a question. Therefore, this study aims to quantify the effects of different assumptions on the \hi \lya intensities of a CME and its driven shocks. This will be accomplished by comparing the results obtained under the following conditions: (1) whether the incident intensity of the solar chromospheric \hi \lya line is uniform or not, (2) whether the influence of the geometric scattering angle is taken into account, and (3) whether the kinetic temperature of neutral hydrogen is assumed to be equal to the electron temperature. 

This work uses a two-temperature (2T, proton and electron temperatures) magneto-hydrodynamic (MHD) model to simulate a fast CME and its driven shock. A brief overview of the 2T MHD model is provided in Section \ref{sec:mhd}. Subsequently, using the parameters obtained from the MHD simulation, we synthesize the CME and shock images in \hi \lya line under different assumptions in Section \ref{sec:synimg}. The synthetic results are then analyzed and compared in Section \ref{sec:results}. Finally, Section \ref{sec:conclusion} shows discussions and conclusions.

\section{Two-temperature MHD simulation}~\label{sec:mhd}
To obtain reasonable physical parameters of a CME for synthesizing the UV \hi \lya intensity map, we make a CME simulation based on an event that occurred on March 7, 2011, originating from active region 11164. The simulation methodology is similar to the one conducted by \citet{JinMeng2017}. In this simulation, the solar corona component is generated using the \alfven~wave damping to heat the corona and considers separate electron and proton temperatures with electron heat conduction \citep{vanderHolst2010,Jin2012ApJ,Sokolov2013,Oran2013ApJ,Jin2013ApJ,JinMeng2017}. The initiation of the CME is achieved by utilizing the Gibson-Low (GL) analytical flux rope model \citep{Gibson1998ApJ}. This model allows for the initiation of a CME from a state of force imbalance, improving computational efficiency. Further details regarding the simulation can be found in \citet{JinMeng2017}. In our study, we decrease the propagated speed of the CME compared to the simulation conducted by \citet{JinMeng2017}. This adjustment is made to mitigate the impact of the DD effect when synthesizing the \lya image while still ensuring the generation of the CME-driven shock. It is important to emphasize that the aim of this work is to investigate the effect of different assumptions on the \hi \lya intensities of the CME and its driven shock. Therefore, the agreement between the simulated CME events and observational data does not affect the results of this study.

\begin{figure}
    \centering
    \includegraphics[width=1\textwidth]{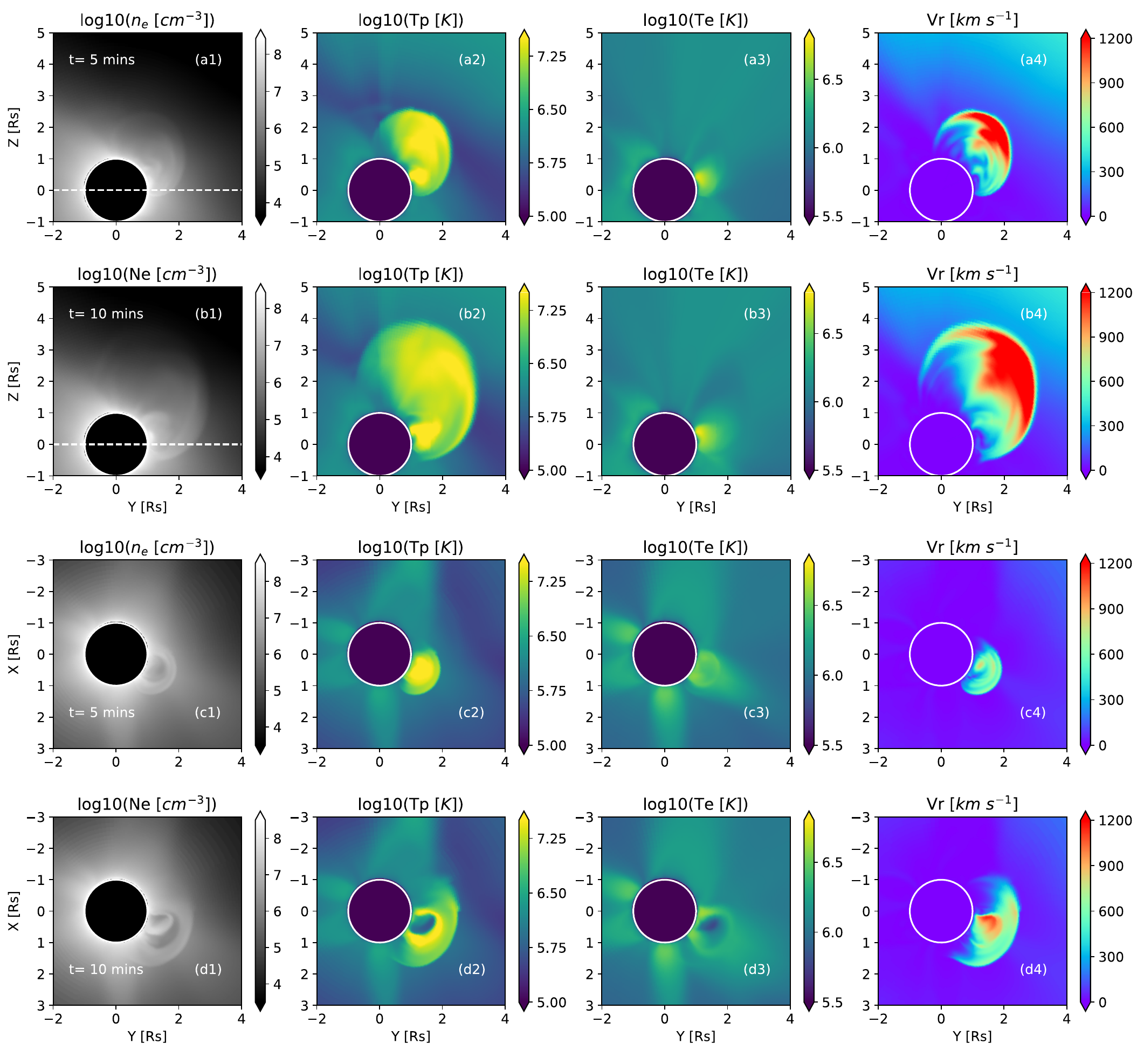}
    \caption{(a) and (b): Parameters of the CME and shock showing the plasma density, proton temperature, electron temperature, and radial speed on the PoS (plane YOZ) from left to right panels at $t= 5$ minutes (first row) and $t= 10$ minutes (second row) after Gibson-Low flux rope implements, respectively. (c) and (d): Parameter cuts of the CME and shock in the plane XOY as seen from the north pole, corresponding to the dashed lines marked in the plane YOZ, as shown in panels (a1) and (b1).}
    \label{fig:mhd_cme}
\end{figure}

\begin{figure}
    \centering
    \includegraphics[width=1\textwidth]{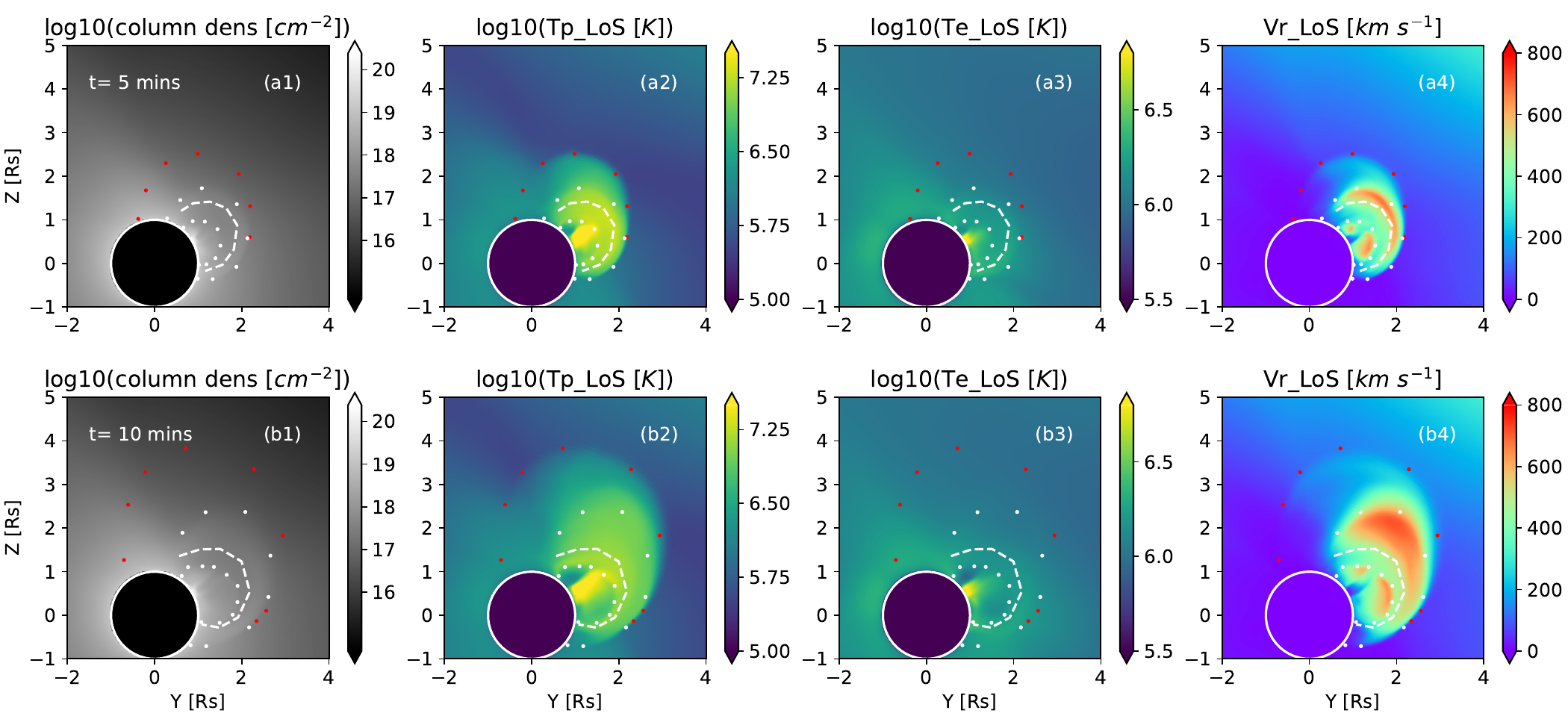}
    \caption{(a) and (b): Parameters of the CME and shock showing the electron column density, integrated proton temperature, electron temperature, and radial velocity along the LoS weighted by the electron number density from left to right panels. CME fronts and cores are marked by white dots, the periphery of CME voids are marked by a white dashed line, and shock fronts are denoted by red dots, which are distinguished from Figure \ref{fig:uv_wl}.}
    \label{fig:mhd_cme2}
\end{figure}

The simulation provides various physical parameters, including the electron number density ($n_e$), the three velocity components along the $x$, $y$, and $z$ axes in Cartesian coordinates, as well as the electron ($T_e$) and proton ($T_p$) temperatures, among others. The data cube dimensions used in this work are interpolated into a uniform grid of $700 \times 300 \times 300$ along the $x$, $y$, and $z$ axes, respectively. The direction of the line of sight (LoS) is aligned with the $x$-axis. The $y$-axis ranges from $-2$ to 4 \rsun, the $z$-axis ranges from $-1$ to 5 \rsun, and the $x$-axis spans from $-7$ to 7 \rsun. Figure \ref{fig:mhd_cme} displays the parameters of the simulation on the plane of the sky (PoS, panels a and b) and in the plane XOY seen from the north pole (panels c and d) at $t=5$ and $t=10$ minutes, respectively. Panels c and d of Figure \ref{fig:mhd_cme} show the parameters in the range of $-3$ to 3 \rsun~along the $x$-axis. It is obvious that the parameters of the CME are asymmetrical in 3D space. The CME front's radial speed ($V_r$) on the PoS is approximately 1200 \kms. In Figure \ref{fig:mhd_cme2}, the first column shows the electron column density ($N_e=\int n_e dz$). The second to fourth columns represent the integrated proton temperature, electron temperature, and radial velocity along the LoS weighted by the electron number density, which can be expressed as $p_{LoS}={\int p~n_e dz}/{\int n_e dz}$, where the parameter $p$ can be $T_p$, $T_e$, and $V_r$. The CME owns a three-part structure, as shown in Figure \ref{fig:mhd_cme2}, and the core is made of a flux rope. CME fronts and cores are marked by white dots, CME voids are marked by a white dashed line, and shock fronts are denoted by red dots. These structures are distinguished from the WL and UV \hi \lya synthetic images (as shown in Figure \ref{fig:uv_wl}) in the next section, based on the physical parameters obtained from the simulation and assuming optical thinness in the corona. Regarding prominences (the core of another category of CME), \citet{Zhao2022} synthesized the \lya emission of an eruptive prominence through the PRODOP code, which takes into account non-local thermodynamic equilibrium radiative transfer processes. However, the discussion of the prominence, considering the optical thickness assumption, is beyond the scope of this study. 

\section{Synthetic images under different assumptions}~\label{sec:synimg}
\subsection{Formation mechanism of \hi \lya lines}
The formation mechanism of \hi \lya lines is a combination of radiative and collisional excitations followed by spontaneous emissions \citep{Gabriel1971}, which can be written as 
    \begin{equation}
        I_{obs}=\int_{LoS}~(j_r+j_c)~dz ~~[\rm photon~s^{-1}~cm^{-2}~sr^{-1}],
        \label{eq:I1}
    \end{equation}
where $j_r$ and $j_c$ are the radiative and collisional emissivities, respectively, from each volume element. The total intensity, $I_{obs}$, is computed by the integration of all elements along the LoS.

The radiative emissivity,$j_r$, due to the resonant scattering of radiation from the lower atmosphere by coronal atoms \citep{Noci1987,Bemporad2018} is
\begin{subequations}
    \begin{align}
        &j_r =\frac{h~B_{12}~\lambda_{0}}{4~\pi}~n_i
        \int_{\Omega}~p(\theta)~d\Omega~\int_{-\infty}^{+\infty}I_{ex}(\lambda^{'}-\delta \lambda,\boldsymbol{n})~
        \Phi(\lambda^{'}-\lambda_0,\boldsymbol{n})~d\lambda^{'}, ~\label{eq:a11} \\
        &p(\theta) =\frac{1}{4\pi}\frac{11+3cos^2\theta}{12},~\label{eq:a2}  \\      
        &\Phi(\lambda^{'}-\lambda_0,\boldsymbol{n}) =\frac{1}{\sigma_{\lambda} (\boldsymbol{n})~\sqrt{2\pi}}~\exp[-\frac{(\lambda^{'}-\lambda_{0})^2}{2\sigma^2_{\lambda} (\boldsymbol{n})}]~~[\rm cm^{-1}],~\label{eq:a3} \\      
        &\sigma_{\lambda}(\boldsymbol{n})=\frac{\lambda_{0}}{c}\sqrt{\frac{k_B~T_{\boldsymbol{n}}}{m_p}}~~[\rm cm], ~\label{eq:a4}
    \end{align}
    ~\label{eq:a0}
\end{subequations}
where $h$ is the Planck constant, $B_{12}$ is the Einstein coefficient for absorption of the considered atom transition, $\lambda_{0}$ is the reference wavelength of the transition in equation \ref{eq:a11}. $n_i$ is the neutral hydrogen number density. $p(\theta)$ is a geometrical function for the scattering process, $\theta$ is the angle between the direction of the incident radiation $\boldsymbol{n}$ and along the LoS toward the observer \citep{Beckers1974,Noci1987,Auchere2005a}. $I_{ex}(\lambda^{'}-\delta \lambda,\boldsymbol{n})$ is the intensity spectrum of the incident chromospheric radiation, $\delta \lambda = \frac{\lambda_{0}}{c}\boldsymbol{u\cdot n}$ is the shift of the incident profile due to the radial velocity, $\boldsymbol{u}$, of the absorbing atoms. $c$ is the light speed. $\Phi(\lambda^{'}-\lambda_0, \boldsymbol{n})$ is the normalized coronal absorption profile along the direction of the incident radiation. It can be assumed as a Gaussian profile with a width, $\sigma_{\lambda}(\boldsymbol{n})$, as shown in equation \ref{eq:a4}. $k_B$ is the Boltzmann constant, and $T_{\boldsymbol{n}}$ is the kinetic temperature along the direction of the incident radiation. In this work, we will assume $T_{\boldsymbol{n}}=T_{p}$, $T_{p}$ is the proton temperature provided by the MHD simulation. It is noted that the kinetic temperature of neutral hydrogen can be considered equal to the proton temperature. This is because when a neutral hydrogen atom forms through the recombination of a proton and an electron, its velocity will be close to that of the proton since the proton carries the majority of the momentum. The assumption can be held when the plasma obeys an isotropic thermal velocity distribution.

The collisional component, $j_c$, of the \lya emission line in the solar corona, is due to the excitation of coronal atoms by collision with free electrons. The formula can be expressed as
\begin{subequations}
    \begin{align}
    &j_c=\frac{1}{4\pi}~n_e~n_i~q_{coll},~\label{eq:b1}\\
    &q_{coll}=2.73\times~10^{-15}~T_e^{\frac{1}{2}}(E_{12})^{-1}~f_{12}~\bar{g}~exp^{-\frac{E_{12}}{k_B T_e}}~~[\rm cm^{-3} s^{-1}],\\
    &n_i=0.833~n_e~R_{HI}(T_e),~\label{eq:b3}
    \end{align}
\end{subequations}
where $n_e$ is the electron number density. $q_{coll}$ is the collisional coefficient, $T_e$ is the electron temperature, $E_{12}$ is the transition energy, $f_{12}$ is the transition oscillator strength, $\bar{g}$ is the Gaunt factor. 0.833 is the proton-to-electron density ratio for a fully ionized gas with a He abundance of 10\%. $R_{HI}(T_e)$ is the elemental ionization fraction as a function of the electron temperature under the assumption of ionization equilibrium.

\subsection{\lya solar disk intensity}
\begin{figure}
    \centering
    \includegraphics[width=1\textwidth]{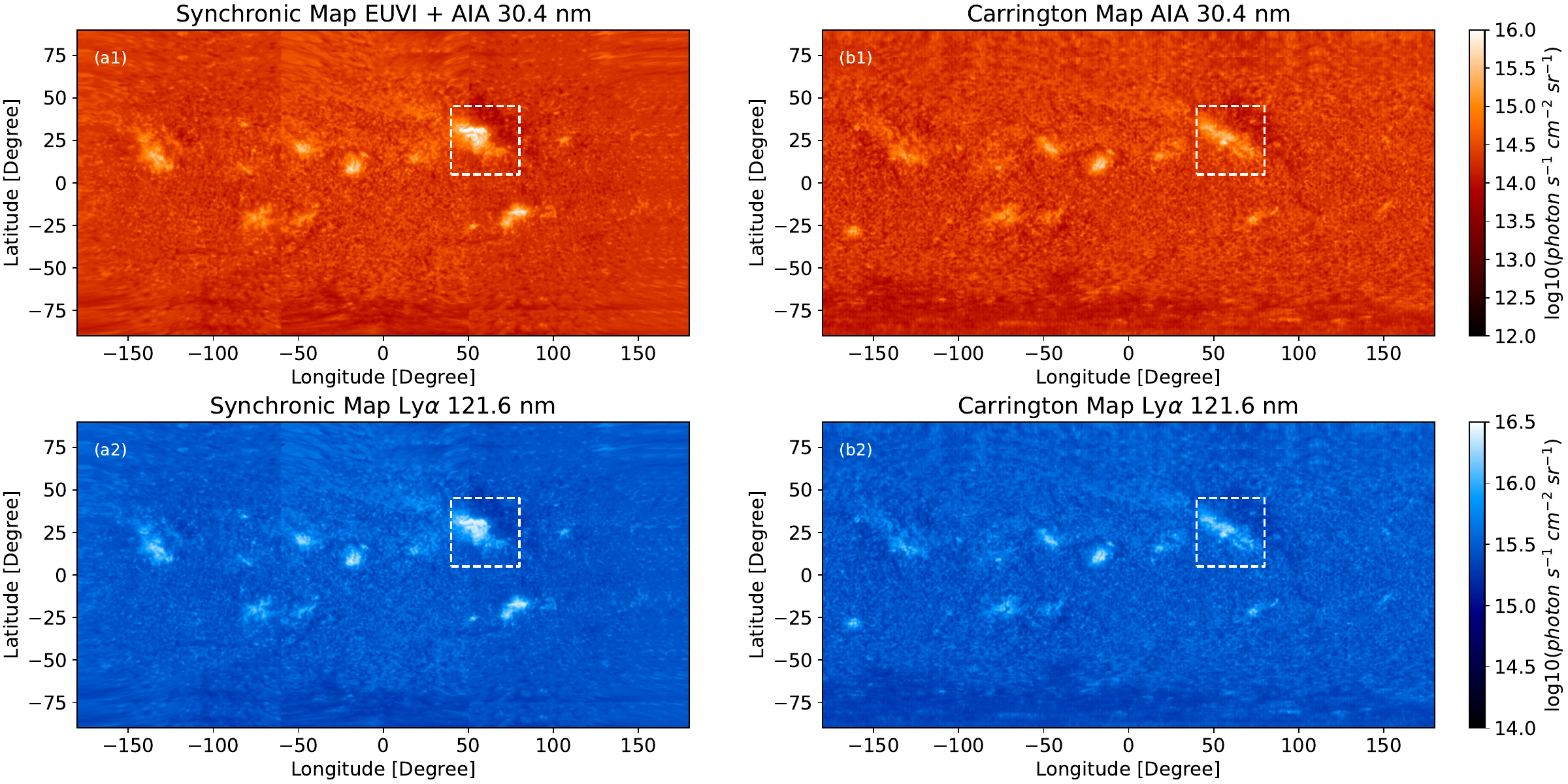}
    \caption{The \heii 30.4 nm and \hi \lya 121.6 nm solar disk intensities in synchronic and Carrington maps. (a1): The synchronic map in 30.4 nm was constructed by the SDO/AIA and STEREO/EUVI images at 20:06 UT on March 7, 2011. (a2): The \hi \lya synchronic map is rebuilt based on the correlation between the \heii 30.4 nm and \hi 121.6 nm in \lya lines. (b1-b2): Carrington maps in \heii 30.4 nm and \hi \lya 121.6 nm lines reconstructed by the SDO AIA 30.4 nm images. White boxes denote the source region of the eruptive CME. The missing data above $+78^{\circ}$ and below $-78^{\circ}$ latitude has been replaced by data in the latitude $\pm [66^{\circ},78^{\circ}]$.}
    \label{fig:disk_iten}
\end{figure}

To generate synthetic UV \hi \lya images of the modelled CME and shock, the intensity spectrum, $I_{ex}(\lambda^{'}-\delta \lambda,\boldsymbol{n})$, of the \lya radiative component is directly influenced by the chromospheric source radiation. $I_{ex}$ can be separated into two terms, including the normalized chromospheric line profile $f(\lambda^{'}-\delta \lambda)$ and chromospheric disk intensity $I_{disk}(\boldsymbol{n})$. In this work, we assume that $f(\lambda^{'}-\delta \lambda)$ is a unique multi-Gaussian profile proposed by \citet{Auchere2005a}, which also has been used to estimate the effect of the non-uniform solar chromospheric \lya radiation on determining the coronal \hi outflow velocity \citep{Dolei2019}. Concerning the \lya solar disk intensity, we build a synchronic map of the \lya intensity based on the correlation between the \heii 30.4 nm and \hi \lya 121.6 nm intensities by combining the images of the Atmospheric Imaging Assembly (AIA) on board the Solar Dynamics Observatory (SDO) and the extreme ultraviolet imager (EUVI) on board the Heliospheric Imager on board the Solar TErrestrial RElations Observatory (STEREO) at the same time, as shown in Figure \ref{fig:disk_iten} (a). The correlation function \citep{Auchere2005a} is 
\begin{equation}
    I_{121.6}=436~[1-0.955\exp(-0.0203~I_{30.4})],
\end{equation}
where $I_{121.6}$ and $I_{30.4}$ are in units of [$\rm W~m^{-2}~sr^{-1}$]. \citet{Gordino2022} found that the empirical relation between the intensities of the \hi 121.6 nm and \heii 30.4 nm \lya lines remains stable over time. However, the accuracy of this relation depends on the spatial resolution of the observations. The contamination of \Sixi 30.32 nm line near the \heii 30.4 nm and other lines have been subtracted by the differential emission measure technique \citep{Cheung2015, SuY2018} in the AIA 30.4 nm image, similar to the method used in \citet{Gordino2022}. Additionally, we conduct a cross-radiometric calibration by analyzing the overlapping region between the AIA and EUVI 30.4 nm images on the solar disk. We assume that the contamination of the EUVI 30.4 nm image in the active region and quiet Sun is equal to the mean value of that in the AIA 30.4 nm image in the active region and quiet Sun. We also build a Carrington map (a diachronic map created by stitching observations made over one Carrington cycle) in the \heii 30.4 nm from the AIA data with the subtraction of the \Sixi contamination and then generate the \hi \lya Carrington map (Figure \ref{fig:disk_iten}, b2). The mean value of the disk intensity in the \hi \lya line for the synchronic map is $3.20\times 10^{15}$ $\rm photon~s^{-1}~cm^{-2}~sr^{-1}$, and for the Carrington map is $3.15\times 10^{15}$ $\rm photon~s^{-1}~cm^{-2}~sr^{-1}$. The mean value of the synchronic map is used as the value of the uniform map. In the following section, these maps are then used to determine the difference in \lya intensity between the inputs of the Carrington, uniform, and synchronic maps. The source region of the eruption is marked by a white box at 20:06 UT on March 7, 2011. The missing data above $+78^{\circ}$ and below $-78^{\circ}$ latitude has been replaced by data in the latitude $\pm [66^{\circ},78^{\circ}]$ in Figure \ref{fig:disk_iten}. 

\subsection{Synthetic images}
\begin{table}
    \caption{Different assumptions for the 2D synthetic \lya intensity. Column 1: Group number. Column 2: $I_{disk}(\boldsymbol{n})$ is the disk intensity of the incident chromospheric radiation. The value of the uniform map is equal to the mean value of the synchronic map. Column 3: The geometric function of the scattering process. We consider the geometric function according to Equation \ref{eq:a2} and ignore this process when assuming $p(\theta)=1/4\pi$. Column 4: $T_{\boldsymbol{n}}$ is the kinetic temperature in Equation \ref{eq:a4}.}
    \label{tab:my_label}
        \begin{tabular}{cccc}
        \hline
            1   & 2 & 3 & 4 \\
            No. & $I_{disk}(\boldsymbol{n})$ & $p(\theta)$ & $T_{\boldsymbol{n}}$ \\ \hline
            Group 1 & synchronic map & Equation\ref{eq:a2} & $T_{\boldsymbol{n}}=T_p$ \\
            Group 2 & Carrington map & Equation\ref{eq:a2} & $T_{\boldsymbol{n}}=T_p$ \\
            Group 3 & uniform map & Equation\ref{eq:a2} & $T_{\boldsymbol{n}}=T_p$ \\
            Group 4 & synchronic map & $p(\theta)=\frac{1}{4\pi}$ & $T_{\boldsymbol{n}}=T_p$ \\
            Group 5 & synchronic map & Equation\ref{eq:a2} & $T_{\boldsymbol{n}}=T_e$ \\
            Group 6 & uniform map & $p(\theta)=\frac{1}{4\pi}$ & $T_{\boldsymbol{n}}=T_e$ \\ \hline
        \end{tabular}
    
\end{table}

\begin{figure}[htb]
    \centering
    \includegraphics[width=1\textwidth]{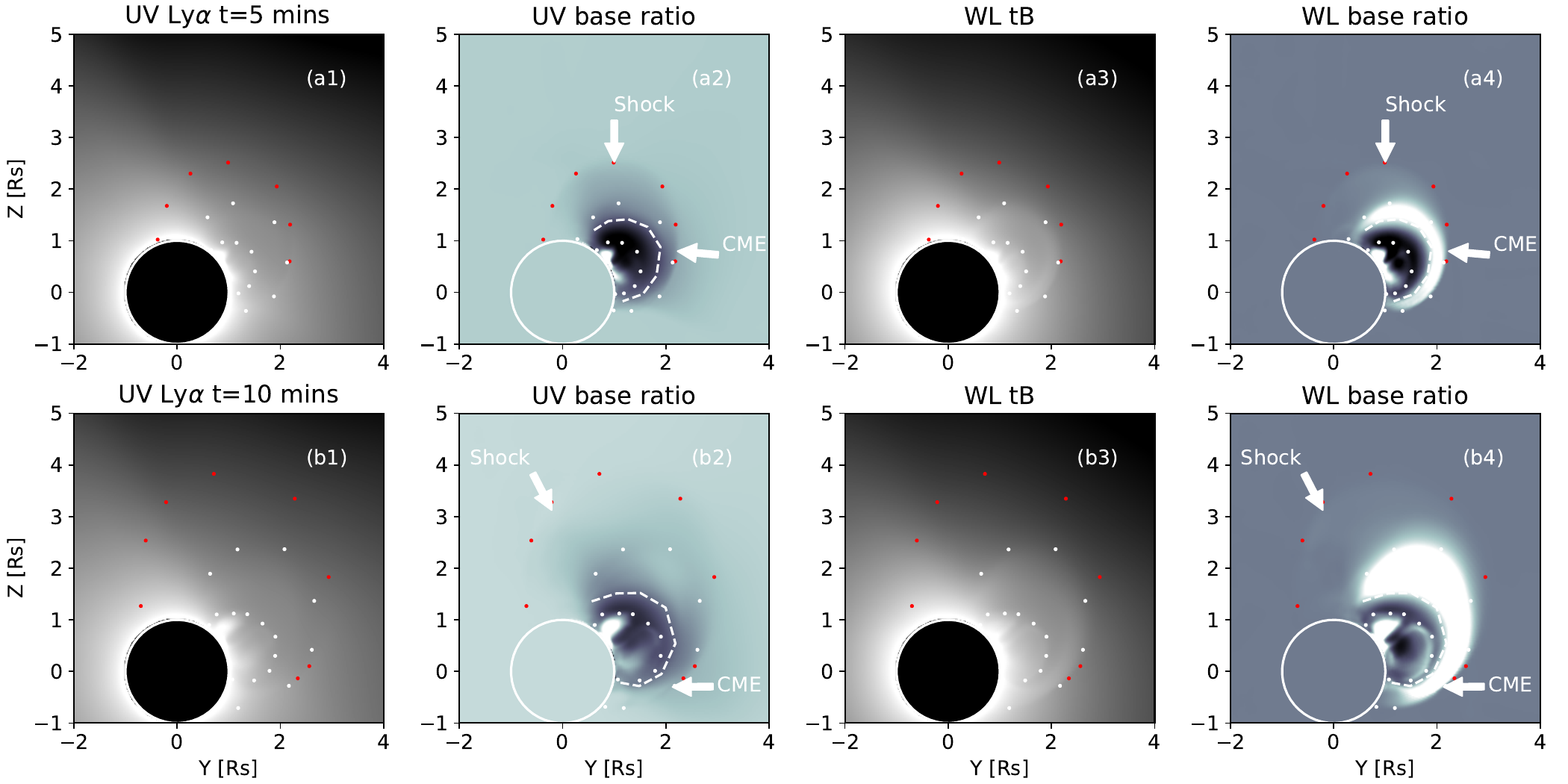}
    \caption{Composited UV \hi \lya (Group 1) and WL images at different simulated times. Panels in columns (1-2) denote the UV \lya original and base-ratio images. Panels in columns (3-4) represent the WL total brightness (tB) and base-ratio images. CME fronts and cores are marked by white dots, the periphery of CME voids are marked by a white dashed line, and shock fronts are denoted by red dots.}
    \label{fig:uv_wl}
\end{figure}
Based on Equations \ref{eq:a11} and \ref{eq:b1} and the physical parameters produced by the MHD simulation, we can calculate the \hi \lya emissivity of each plasma in the data cube and obtain the 2D synthetic \lya image with the intensity integration along the LoS. The composited \lya images are divided into six groups depending on the different assumptions of three physical parameters in the \lya radiative component, including the disk intensity $I_{disk}(\boldsymbol{n})$, the geometric scattering function $p(\theta)$, and the kinetic temperature $T_{\boldsymbol{n}}$, as shown in Table \ref{tab:my_label}. Group 1 will be a reference group in what follows. The assumptions of Group 6 have been applied frequently to estimate the CME electron temperatures \citep{Susino2016, Bemporad2018, Ying2020}. Then, we will compare the \lya intensity difference of these groups to the reference group in the next section.

Examples of the synthetic UV \lya images under assumptions of Group 1 are shown in Figure \ref{fig:uv_wl}. From left to right in Figure \ref{fig:uv_wl}, these panels are UV original, UV base-ratio (normalized by the coronal background image), WL total Brightness (tB), and WL base-ratio images, respectively. WL synthetic image is generated by integrating of the electron number density ($n_e$) along the LoS and multiplying it by geometrical functions that depend solely on the heliocentric distance \citep{Billings1966}. CME fronts and cores are marked by white dots, CME voids are marked by a white dashed line, and shock fronts are denoted by red dots. Similar to the UV images synthesized by \citet{Bemporad2018}, CME fronts are darker than the coronal background. The shock regions are indistinguishable in UV \lya original images and can only be recognizable in base-ratio images, as shown in Figure \ref{fig:uv_wl}. Previous studies have observed that the \lya intensity of the shock decreases by $\sim 30\%$ \citep{Ciaravella2005} and manifests as a dimming region with the shock transition \citep{Bemporad2010} in the UVCS slit-spectroscopic observations. In this simulation, the \lya intensity of the synthetic shock wave is approximately 20-30\% lower than that of the corona background. However, as the shock evolves and the number density decreases (Figure \ref{fig:uv_wl}, b4), the \lya intensity difference between the shock and the background gradually diminishes (Figure \ref{fig:uv_wl}, b2).

\section{\lya intensity uncertainties under different assumptions}~\label{sec:results}
In this section, we derive the relative uncertainties of the \hi \lya intensity by comparing the 2D synthetic intensities of different groups (2-6) with respect to Group 1 to quantify the effect of these assumptions on the \lya total radiation of the CME and shock.

\subsection{Uncertainties from different solar disk maps}
\label{sec:disk_diff}
\begin{figure}[htb]
    \centering
    \includegraphics[width=0.7\textwidth]{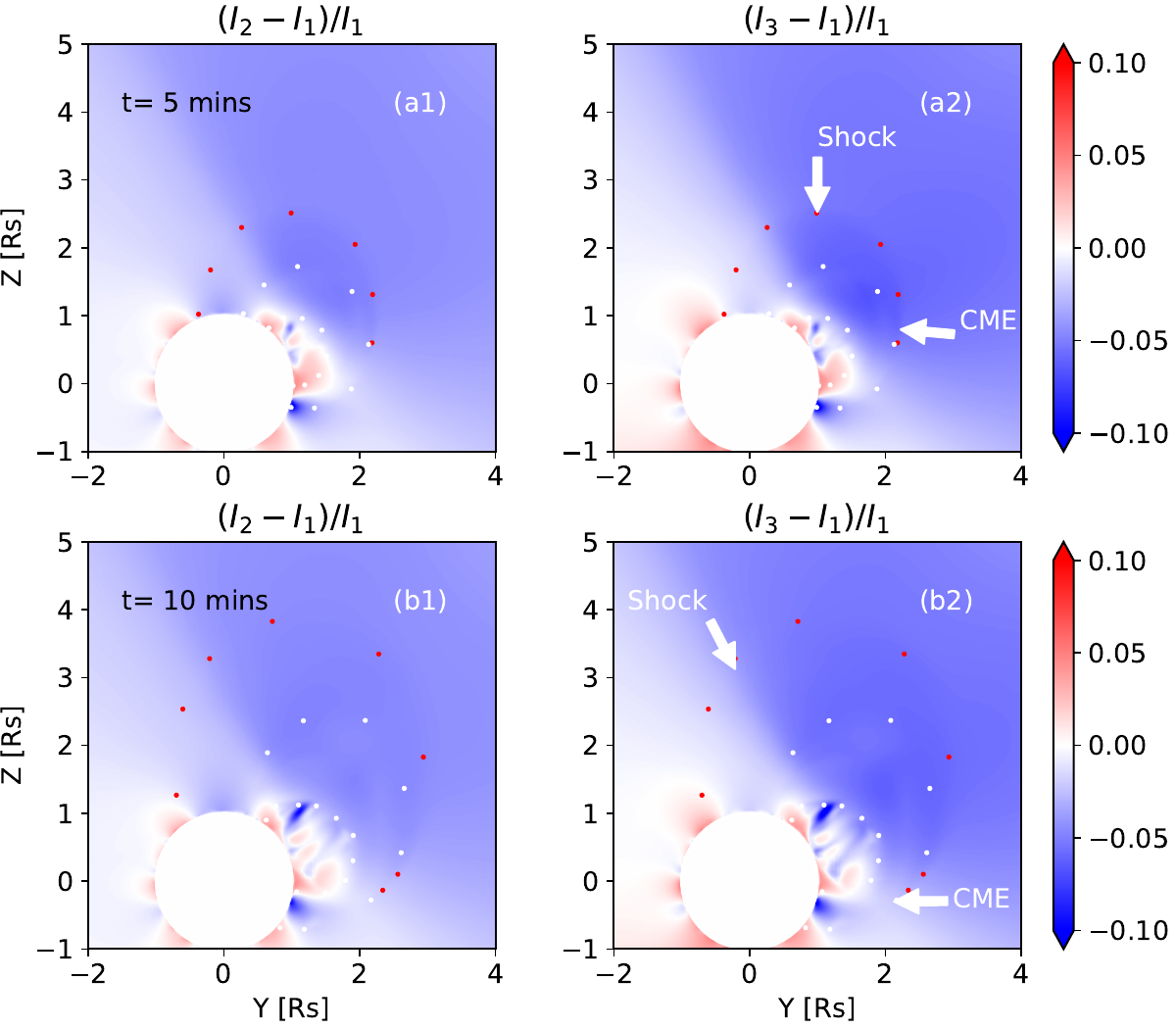}
    \caption{Relative comparisons of \lya intensities calculated by the Carrington (Group 2, left panels) and uniform (Group 3, right panels) maps with respect to the synchronic (Group 1) map at $t= 5$ minutes (top panels) and $t= 10$ minutes (bottom panels), respectively. Arrows and dots with different colors have the same locations as those in Figure \ref{fig:uv_wl}. }
    \label{fig:disk_diff}
\end{figure}

Figure \ref{fig:disk_diff} illustrates the relative comparisons of the corona \lya intensities derived with the Carrington map (Group 2) and uniform map (Group 3) with respect to the synchronic map (Group 1). The only differing parameter among these three groups, when calculating the \lya intensity, is $I_{disk}(\boldsymbol{n})$, as depicted in Table \ref{tab:my_label}. In Figures \ref{fig:disk_diff} (a1) and (b1), it can be observed that the calculated \lya intensities using Carrington map and synchronic map as inputs for $I_{disk}(\boldsymbol{n})$ exhibit small differences in most regions, with relative underestimate uncertainties remaining within 6\%. A partial core is overestimated ($<$4\%), whereas the other parts are primarily underestimated. The largest discrepancies between Group 2 and Group 1 are observed in the CME core, with a maximum underestimate uncertainty of approximately 10\% (in panel b1). This region has relatively lower electron and proton temperatures and higher electron number densities compared to other regions. It is possible that the observed increase in the \lya intensity of Group 1 is attributed to the influence of a flare present in the synchronic map within the source region (indicated by a white box in Figure \ref{fig:disk_iten}). However, it should be noted that the impact of the flare is relatively minor or insignificant in comparison. Comparing Group 3 to Group 2,  it is evident that Group 3 exhibits an overall increase in relative underestimated uncertainties, with a maximum of 13\%. Both the Carrington map and uniform map, when used as inputs for $I_{disk}(\boldsymbol{n})$, underestimate the \lya intensities of the CME and shock, except for part of the CME core. However, the results from the Carrington map align more closely with those from the synchronic map. Therefore, when estimating the electron temperature of the CME using the DD method in the absence of synchronic map observations, we recommend utilizing the Carrington map as the input for \lya $I_{disk}(\boldsymbol{n})$, which conveniently coincides with the data provided by LST/SDI observations, to obtain the theoretical \lya intensity. In comparison to the CME, the impact of a uniform distribution of incident radiation on the \lya intensity of the shock is relatively insignificant.

\subsection{Uncertainties from the geometric scattering process}
\begin{figure}[htb]
    \centering
    \includegraphics[width=1\textwidth]{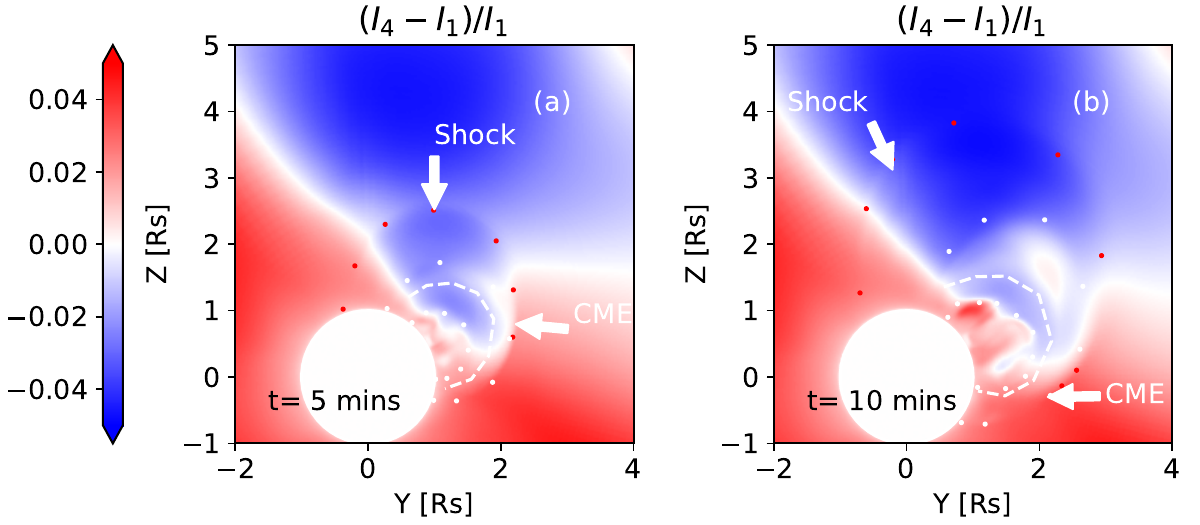}
    \caption{Relative variations of \lya intensities calculated by assumptions of Group 4 with respect to Group 1 at $t= 5$ minutes and $t= 10$ minutes, respectively. White arrows and dots (red and white) have the same locations as those in Figure \ref{fig:uv_wl}.}
    \label{fig:geo_diff}
\end{figure}

In this subsection, We will assess the impact of neglecting the geometric scattering angle on the theoretical calculation of the \lya intensity. Figure \ref{fig:geo_diff} illustrates the relative variations in \hi \lya intensity between the hypothetical calculations of Group 4 (ignoring geometric scattering function) and Group 1 at $t=5$ and $t=10$ minutes, respectively. In Figure \ref{fig:geo_diff}, we can find that the CME core is overestimated in the evolution, and the maximum relative uncertainty is less than 5\%. For the CME void and front, both overestimates and underestimates are present, but the underestimation is dominant. Regarding the shock, most parts are underestimated. Then, with the evolution of the shock and the decrease in number density, the underestimation uncertainty increases and becomes comparable to that of the background, as depicted in Figure \ref{fig:geo_diff} (b). It could be attributed to the lower number density of shocks, where the \lya intensity is more influenced by the coronal background after the LoS integration. On the other hand, the absolute value of the maximum uncertainty between the two groups does not exceed 5\%.

\subsection{Uncertainties from temperature assumptions}
\label{sec:temp_diff}

Due to the absence of spectroscopic or in-situ proton temperature observations in the corona, it is common to assume that the proton temperature is equal to the electron temperature when calculating the \hi \lya theoretical intensity \citep{Bemporad2018}. This simplified assumption enables us to diagnose the CME electron temperature using the DD approach. However, it raises the question of how much uncertainty is introduced in calculating the \lya theoretical intensity under this assumption. In this subsection, we will quantify the impact of the assumption of equal proton and electron temperatures on the uncertainty associated with the computation of the \lya theoretical intensity.
\begin{figure}
    \centering
    \includegraphics[width=1\textwidth]{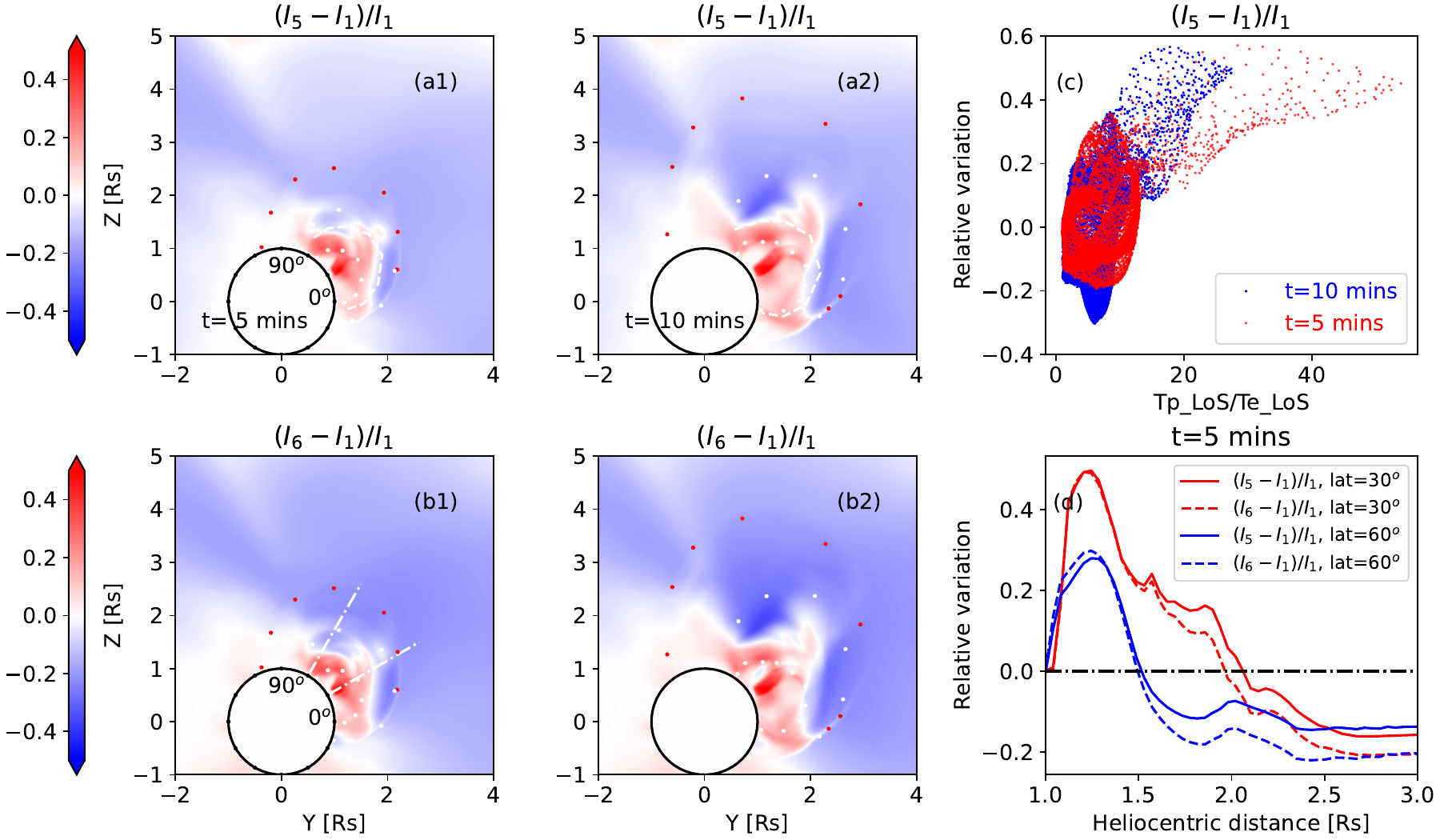}
    \caption{Relative variations of \lya intensities calculated by assumptions of Groups 5 and 6 with respect to Group 1 at $t= 5$ minutes and $t= 10$ minutes, respectively. (a1-a2): The relative variations of Group 5 to Group 1 for the whole CME and shock. (b1-b2): The relative variation distributions of Group 6 to Group 1. Dashed-dotted lines in (b1) are along radial directions at latitudes of 30$^{\circ}$ and 60$^{\circ}$. Red, white dots, and white dashed curves in panels (a) and (b) have the same locations as those in Figure \ref{fig:uv_wl}. (c): Scatter plot of the relative variation (Group 5) vs. the LoS proton-to-electron temperature ratio. Symbols include all points inside the CME and shock both at $t=5$ (red) and $t=10$ (blue) minutes. (d): Relative variations of Groups 5 (solid line) and 6 (dashed line) at $t=5$ minutes along the radial directions at latitude of 30$^{\circ}$ and 60$^{\circ}$. A horizontal line denotes a constant value of zero.}
    \label{fig:temp_diff}
\end{figure}

Figure \ref{fig:temp_diff} (a1) and (a2) illustrate the comparisons between Group 5 and Group 1 at the 5th and 10th minute of the simulation, respectively. In Group 5, we set the kinetic temperature equal to the electron temperature, while Group 1, serving as the reference group, assumes the kinetic temperature equal to the proton temperature. The relative uncertainty distribution in this group is more complex than in other groups, and it exhibits both overestimation and underestimation of the \lya intensity of the internal CME structure. We find that temperature assumptions significantly impact the theoretical \lya intensity. The CME and void are always overestimated when compared to the reference group. The CME core demonstrates a maximum relative uncertainty exceeding 50\%, while the void exhibits a relative uncertainty of no more than 35\%. Regarding the CME front, both overestimation and underestimation are observed, with a relative uncertainty less than 35\%. Compared to the CME, the temperature assumptions have a relatively small impact on the shock, and most part of the computed \lya intensity is generally underestimated, with a relative uncertainty of less than 20\%. Panels (b1) and (b2) depict the relative uncertainties between Group 6 (common assumptions for diagnosing the electron temperatures of CMEs) and the reference group. The uncertainty distribution is a result mixed by these three parameters simultaneously, and the temperature has the most significant influence than the other two parameters ($I_{disk}$ and $p(\theta)$) via directly comparing panels (a) and (b) in Figure \ref{fig:temp_diff}. We make two slits along the radial directions at latitudes of $30^{\circ}$ and $60^{\circ}$ (dashed-dotted lines, as shown in Figure \ref{fig:temp_diff}, b1), and the relative uncertainties are plotted in the panel (d) of Figure \ref{fig:temp_diff}; solid lines denote the \lya intensity uncertainty from Group 5, while dashed lines are from Group 6. Compared to Group 5, Group 6 exhibits decreasing relative uncertainties for the most part, indicating a slight reduction in overestimation and an increase in underestimation.

Figure \ref{fig:temp_diff} (c) presents a scatter plot of the relative variation (from Group 5 at $t=5$ and $t=10$ minutes) of the \lya intensity versus the ratio of proton-to-electron temperature for all data points within the CME and shock. The proton temperature and electron temperature are obtained as the line-of-sight-averaged temperatures weighted by the electron number density (Figure \ref{fig:mhd_cme2}, a2-a3 and b2-b3). The temperature ratio does not show a good correlation with the relative uncertainty and is less than 15 for most plasma in this simulation. Therefore, it is exceedingly challenging to evaluate directly the uncertainty caused by the equal-temperature assumption through the ratio of proton-to-electron temperatures.
\begin{figure}
    \centering
    \includegraphics[width=1\textwidth]{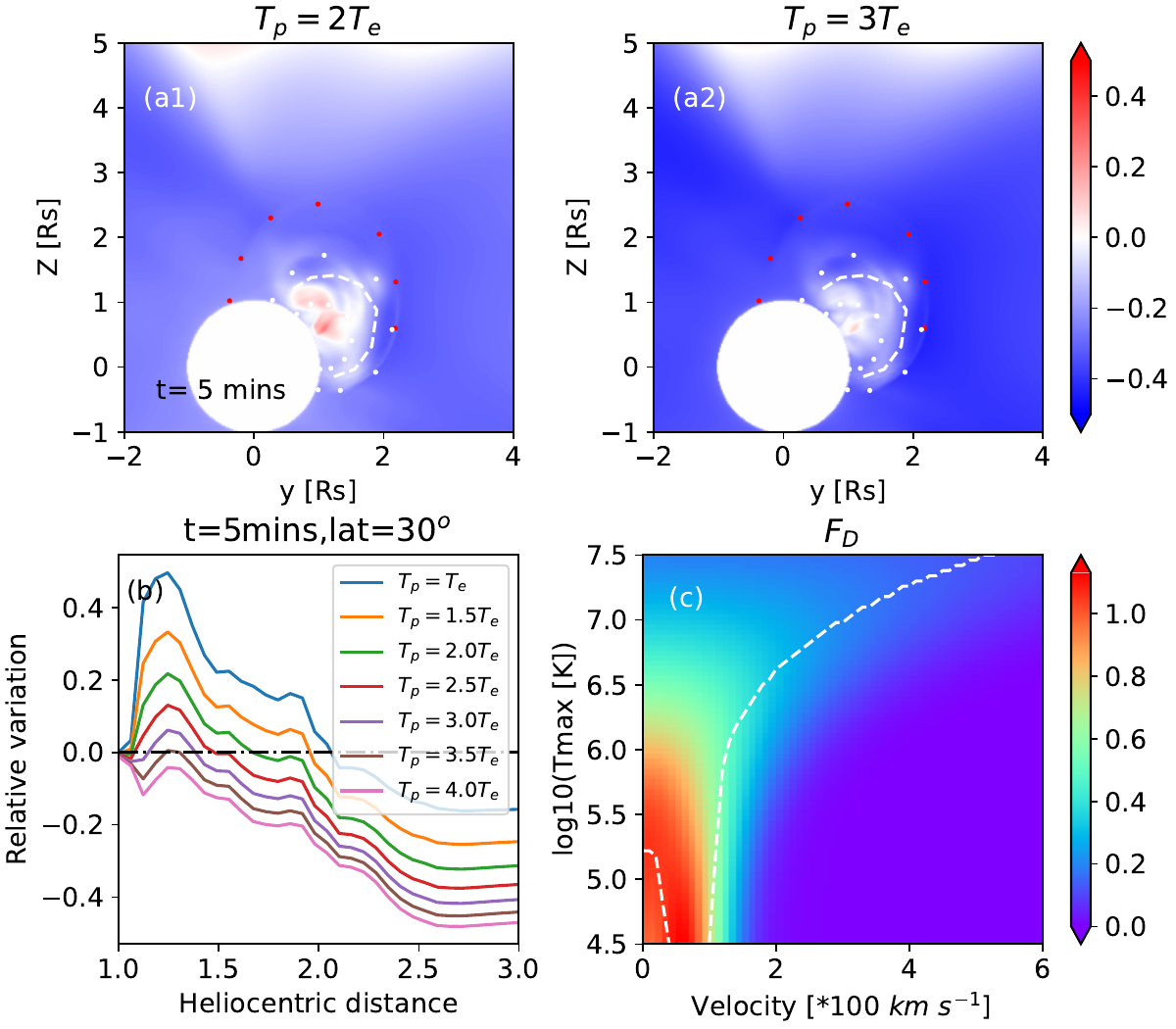}
    \caption{(a1)-(a2): Relative variations of \lya intensities calculated by assuming $T_p=2T_e$ and $T_p=3T_e$, respectively. (b): Relative variations at $t=5$ minutes along the radial direction at the latitude of 30$^{\circ}$ (dashed-dotted line in Figure \ref{fig:temp_diff}, b1) assuming the proton temperature to be equal to various multiples of the electron temperature. (c) Doppler dimming (DD) factor distribution as a function of the radial velocity and temperature. A white dashed line marks the critical temperature corresponding to the maximum DD factor at different speeds.}
    \label{fig:temp_diff_ratio}
\end{figure}
The reason is that the variation of the kinetic temperature ($T_{\boldsymbol{n}}$) first influences the coronal normalized absorption profile (Equations \ref{eq:a3} and \ref{eq:a4}), which is then combined with the normalized incident spectral line profile (subsection \ref{sec:disk_diff}) incident from the solar disk, collectively affecting the \lya intensity. We can define it as a DD factor $F_D=\int f(\lambda^{'}-\delta\lambda)\Phi(\lambda^{'}-\lambda_0,\boldsymbol{n})d\lambda^{'}$, when we separate $I_{ex}(\lambda^{'}-\delta\lambda, \boldsymbol{n})$ (in Equation \ref{eq:a11}) and extract the chromospheric incident intensity $I_{disk}(\boldsymbol{n})$ from the integral equation. Thus, the change of the DD factor (further the \lya intensity) will be influenced by not only the kinetic temperature but also the radial velocity of the plasma. Figure \ref{fig:temp_diff_ratio} (c) illustrates the distribution of the DD factor with respect to the radial velocity and kinetic temperature. A dashed line in the figure shows the evolution of the corresponding temperature with velocity when the DD factor is maximized. 

Considering that the proton temperature is consistently higher than the electron temperature, we further explore the possibility of adjusting the multiples of the electron temperature, instead of assuming equal proton and electron temperatures, to mitigate the uncertainty in the computed \lya theoretical intensity. The results are shown in Figure \ref{fig:temp_diff_ratio}. Panels (a1) and (a2) display relative uncertainties of the \lya intensity calculated by assuming $T_p=2T_e$ and $T_p=3T_e$, respectively. Panel (b) indicates the relative uncertainties with different proton-to-electron temperature ratios at $t=5$ minutes along the radial direction at the latitude of 30$^\circ$. Compared to Group 5 (Figure \ref{fig:temp_diff}, a1), we observe that an appropriate increase in the proton-to-electron temperature ratio leads to a reduction in the relative uncertainty of the CME core and void. Thus, we suggest that when applying the DD method to diagnose the electron temperature in the CME core (made of the flux rope), the theoretical intensity can be computed by assuming the proton temperature to be a few times greater than the electron temperature, such as 2-3 times the electron temperature. But for the CME front and the shock with higher speeds, the \lya theoretical intensity calculated under the assumption of the equal proton and electron temperature is better than that computed by the higher proton-to-electron temperature ratio. The increasing underestimation in the CME front and the shock (Figure \ref{fig:temp_diff_ratio}, b) can be attributed to the significant influence of the coronal background in those regions, where the DD effect is pronounced due to the high velocity of the shock and CME front. This underestimation effect becomes evident when comparing the backgrounds in Figure \ref{fig:temp_diff_ratio} (a1-a2) with Figure \ref{fig:temp_diff} (a1).


\section{Discussion and Conclusions}~\label{sec:conclusion}
In this work, we model a fast CME with a three-part structure and its driven shock through a 2T numerical simulation. Based on the simulated physical parameters,  we synthesize 2D UV \hi \lya images. We conduct analyses for six groups of different assumptions, with Group 1 serving as the reference group. The assumptions in Group 1 include using a synchronic map as the input for the solar disk intensity ($I_{disk}(\boldsymbol{n})$), considering the geometric scattering process ($p(\theta)$), and assuming the kinetic temperature of hydrogen atoms to be equal to the proton temperature ($T_{\boldsymbol{n}}=T_p$). By comparing the intensities with Group 1, we assess the impact of these different assumptions on the theoretical \lya intensities of the CME and its driven shock. The relative uncertainties in various structures for different groups are summarized in Table \ref{tab:02}. The main conclusions are as follows:

1. $I_{disk}(\boldsymbol{n})$: the relative comparisons of the \lya intensities between the Carrington map (Group 2) and the synchronic map (Group 1) exhibit small differences in most regions, with relative underestimate uncertainties remaining within 6\%. The maximum underestimated uncertainty is around 10\% in the CME core, with low temperature and high density. A part of the CME core is overestimated but the relative uncertainty is less than 4\%. The underestimation of the \lya intensity, when using the uniform map as the input for $I_{\text{disk}}(\boldsymbol{n})$ with a maximum relative uncertainty of approximately 13\%, is greater compared to the results obtained from the Carrington map. Therefore, it is preferable to utilize the \lya Carrington map as the input for $I_{{disk}}(\boldsymbol{n})$ provided by the LST/SDI \citep[More details can refer to ][]{Lishuting2023} when calculating the theoretical \lya intensity of the CME and shock in the absence of synchronic map observations.

2. $p(\theta)$: ignoring the geometric scattering process has a lesser impact on the results of the \hi \lya intensity for the CME and shock. The absolute value of the maximum relative uncertainty does not exceed 5\%. For the most part, the \lya intensity is underestimated, and the CME core is overestimated. With the evolution of the shock, the underestimation uncertainty increases and becomes comparable to that of the background. This could be attributed to the fact that the lower number density of the shock is more susceptible to the influence of the coronal background. It should be noted that the geometric scattering function is valid primarily in weak magnetic fields ($B < 70$ G) \citep{Beckers1974}, which is generally applicable to most of the quiet corona. However, when plasma is located closer to an active region, particularly near the flare site where the magnetic field strength increases, the geometric scattering function will be changed. Such as the magnetic strength of a small C2.0 flare measured by \citet{Landi2021ApJ} can reach up to 500 G. On the other hand, \citet{Zhao2021ApJ} conducted a study about the impact of the magnetic field on the linear polarization of the \hi \lya radiation and found that the variation of the magnetic field will not affect the \lya intensity, but the linear polarization. Therefore, the estimated relative uncertainty from the geometric scattering assumption in this paper can be valid.

3. $T_{\boldsymbol{n}}$: The reference \lya intensity (Group 1) is calculated by assuming that the hydrogen kinetic temperature is equal to the proton temperature, as the simulation only provides the proton and electron temperatures. We find that the temperature assumptions have a significant impact on the theoretical \lya intensity. The distribution of the relative uncertainty is more complex than other groups. The CME core and void are overestimated. In the CME core, the maximum value of the relative uncertainty exceeds 50\% between the electron (Group 5) and proton (Group 1) temperature assumptions, while the relative uncertainty for the void remains below 35\%. We also find that assuming the proton temperature to be a multiple of the electron temperature, such as 2-3 times the electron temperature, can reduce the uncertainty in the CME core and void. In the CME front, we observe both overestimation and underestimation with uncertainties of less than 35\%. 
In contrast, the temperature assumption has a smaller impact on the shock than the CME, resulting in a generally underestimated uncertainty of approximately 20\%. For the CME front and the shock region, assuming equal proton and electron temperatures yields better results compared to those calculated using a higher proton-to-electron temperature ratio.

\begin{table}[]
\caption{Relative uncertainties for different groups shown in Table \ref{tab:my_label}. Column 1: Hypothetical parameters different with Group 1. Column 2: Group number. Column 3: Relative uncertainties in different structures.}
\label{tab:02}
\renewcommand{\arraystretch}{2.4}
\begin{tabular}{ccl}
    \hline
    Parameter                                       & No.     & \multicolumn{1}{c}{Relative uncertainties} \\ \hline
    {$I_{disk}(\boldsymbol{n})$} & Group 2 & \resizebox{!}{!}{
    \begin{tabular}[c]{@{}c@{}}CME core: overestimate ($<4\%$) and underestimate ($<10\%$)\\ CME void, front, and shock: underestimate ($<6\%$) \\  
    \end{tabular}} \\   \cline{3-3}                                                                                                 
    & Group 3 & \begin{tabular}[c]{@{}c@{}} CME core: overestimate ($<4\%$) and underestimate ($<13\%$)\\ CME void, front, and shock: underestimate ($<8\%$)\end{tabular} \\   \cline{3-3}                                                             
    
    $p(\theta)$        & Group 4 & \begin{tabular}[c]{@{}c@{}}CME core: overestimate ($<5\%$)\\ CME front and void: overestimate ($<3\%$) and underestimate ($<2\%$)\\ Shock: overestimate ($<3\%$) and underestimate ($<5\%$) \\ \end{tabular}   \\ [1ex] \cline{3-3}    
    
    $T_{\boldsymbol{n}}$            & Group 5  &  \begin{tabular}[c]{@{}c@{}}CME core: overestimate ($<57\%$)\\ CME void: overestimate ($<35\%$) \\ CME front: overestimate ($<20\%$) and underestimate ($<35\%$)\\ Shock: overestimate ($<10\%$) and underestimate ($<20\%$)\end{tabular} \\ [1ex] \hline

\end{tabular}
\end{table}

In this work, the simulated CME owns a very high proton temperature, according to the LoS integrated proton temperature weighted by the electron number density, as shown in Figure \ref{fig:mhd_cme2} (a2) and (b2). The ratio between the LoS proton and electron temperatures is very large. The significantly higher proton temperature compared to the electron temperature in the simulated CME, by more than one order of magnitude, may differ from the actual CME's proton temperature. On the other hand, \citet{Romeo2023ApJ} reported a CME captured by the Parker Solar Probe (PSP) on 2022 September 5. PSP traversed through the CME and measured many physical parameters, including proton and electron temperatures, at a heliocentric distance of $\sim$ 14 \rsun. In this CME, the proton temperature is consistently higher than the electron temperature, with the maximum proton temperature reaching over 11 MK while the electron temperature remains below 1 MK. Besides, \citet{Reeves2011ApJ} constrained the temperature of magnetic flux ropes within the range of 5 MK to 20 MK, indicating the possibility of MFRs attaining flare-like temperatures. These findings support that a CME can possess a high temperature in the core and a significant proton-to-electron temperature ratio in the body. Thus, the uncertainty estimation of this work could be reasonable. In the absence of proton temperature measurements, we recommend assuming that the proton temperature is a multiple of the electron temperature, as we mentioned above. This assumption could help reduce the uncertainty when calculating the \lya theoretical intensity using the DD method to diagnose the electron temperature.

We ignore the effect of the incident chromospheric \lya spectral profile and assume it is a unique multiple-Gaussian line profile. \citet{Capuano2021} had discussed the impact of the chromospheric \lya line profile on determining the coronal \hi atom speed through the DD method. They considered four parameters of the \lya line profile, including line width, reversal depth, asymmetry, and distance of the peaks. The result demonstrates that the variation of the \lya line profile impacts the coronal \hi velocity estimation by about 9-12\%. They suggested that it is acceptable to use a unique shape of the \lya profile over the solar disk that is constant in time.

Furthermore, it is important to note that the calculation of the theoretical \lya intensity in this study is based on the assumption of ionization equilibrium. However, it is known that the corona is generally not in equilibrium. A freeze-in distance exists for atoms and ions, where lighter elements stop evolving earlier than heavier elements \citep{Landi2012ApJ}. An MHD simulation conducted by \citet{Pagano2020} revealed significant non-equilibrium ionization effects in the front of a CME, while the equilibrium ionization assumption remains valid in the core of the CME. Therefore, it becomes necessary to consider the impact of non-equilibrium ionization when calculating the theoretical \lya intensity.

\begin{acks}
We are grateful to Meng Jin for his help in the 2T MHD simulation. We also thank the anonymous referee for suggestions and comments that helped us significantly improve and clarify this paper. This work is supported by the Strategic Priority Research Program of the Chinese Academy of Sciences, Grant No. XDB0560000, National Key R\&D Program of China 2022YFF0503003 (2022YFF0503000), NSFC (grant Nos. 11973012, 11921003, 12203102, 12233012, 12103090), the mobility program (M-0068) of the Sino-German Science Center. This work benefits from the discussions of the ISSI-BJ Team ``Solar eruptions: preparing for the next generation multi-waveband coronagraphs"
\end{acks}


\appendix   
\section{Calculation of the resonant component of the \hi \lya intensity}

\begin{figure}[h]
        \centering
        \includegraphics[width=1\textwidth]{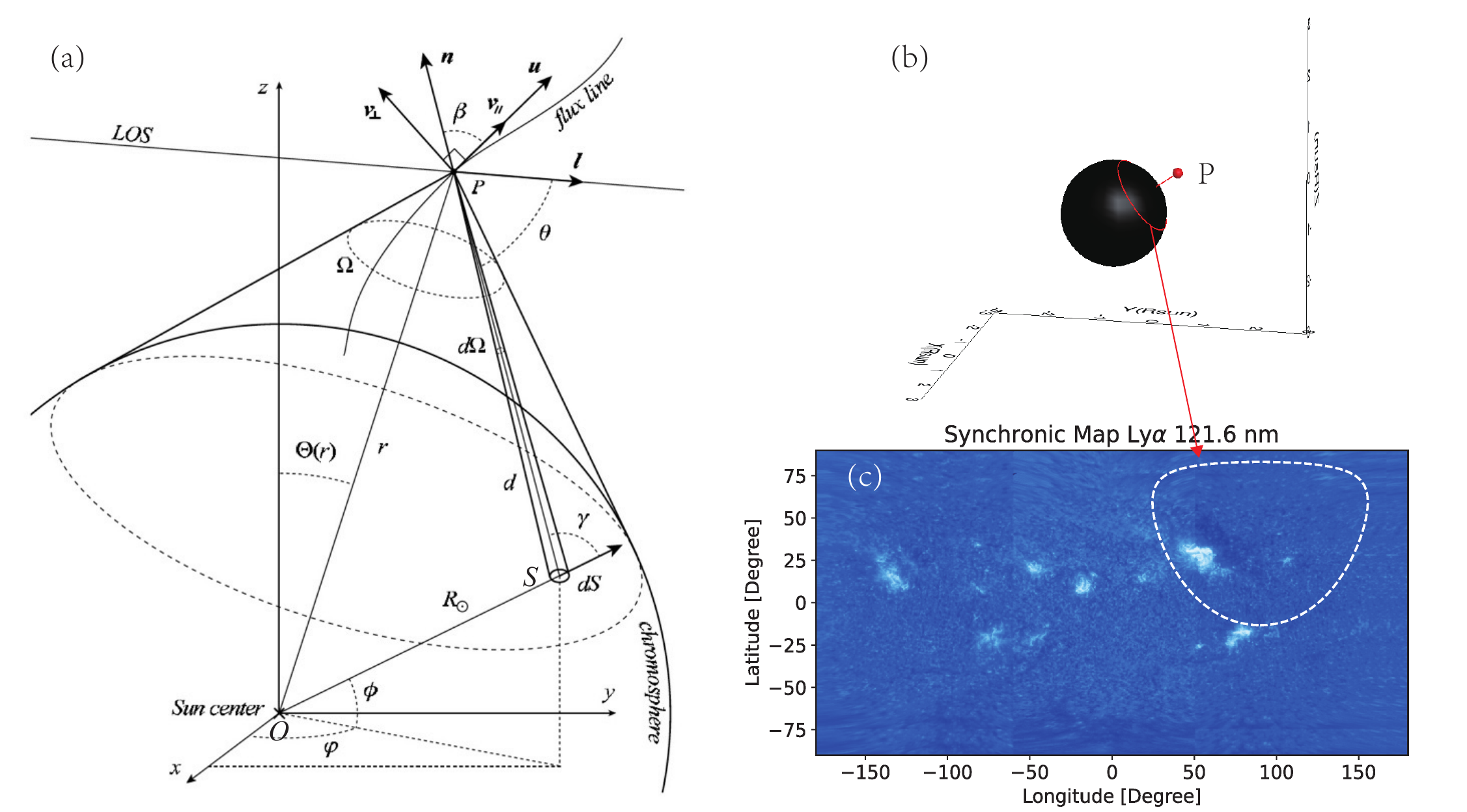}
        \caption{(a) Geometry of the modeling of the resonantly scattered coronal \hi \lya radiation from \citet{Auchere2005a}. (b): The region of incident intensity (red line) on the solar surface (black ball) for a given point P (red ball, $\varphi=90^{\circ}$, $\phi=35^{\circ}$, $r=1.5$ \rsun). (c): The synchronic map in the \lya line of the solar disk, the same as that in Figure \ref{fig:disk_iten} (a2). A white dashed line indicates the region (red circle, panel b) of solar disk incident intensity corresponding to point P after the solar surface unfolding in the Stonyhurst heliographic coordinate.}
        \label{fig:app}
\end{figure}

In this section, we describe calculation details of the resonantly scattered \hi \lya intensity. It mainly refers to \citet{Noci1987}, \citet{Auchere2005a}, \citet{Bemporad2018}, and \citet{Dolei2019}. 

For a given plasma at point $P$ in the corona, its emissivity is contributed by the incident \lya radiation from its corresponding disk region (Figure~\ref{fig:app} a). We can determine its corresponding disk region of the incident \lya radiation, such as the region marked by the dashed line in the unfolding solar disk in the Stonyhurst heliographic coordinate, as shown in Figure~\ref{fig:app} (c). For a grid point S in the disk region (panel a), the direction of $\vec{SP}$ indicates the incident direction $\boldsymbol{n}$, and we can calculate $dS=$\rsun$^{2}\cos{\phi}~d\phi d\varphi$. The solid angle $d\Omega=dS(\cos{\gamma})/d^2$, where $d$ is the modulus of $\vec{SP}$, and $\gamma$ is the angle between $\vec{OS}$ and $\vec{SP}$. The angle $\beta$ is determined from $\boldsymbol{u} \cdot \boldsymbol{n}$, and $\boldsymbol{u}$ is the speed of the plasma at point P. $\theta$ is the geometric scattering angle and can be computed by $\boldsymbol{l} \cdot \boldsymbol{n}$, and $\boldsymbol{l}$ is the direction of the light of sight. In our calculation, $\boldsymbol{l}$ is parallel to the $x$-axis direction. 

As we mentioned in subsection 3.2, $I_{ex}$ can be separated into two terms, and $I_{ex}(\lambda^{'}-\delta \lambda,\boldsymbol{n})={I}_{disk}(\boldsymbol{n})~f(\lambda^{'}-\delta \lambda)$. We assume that the normalized chromospheric line profile is a unique multi-Gaussian profile proposed by \citet{Auchere2005a}, which is
\begin{subequations}
    \begin{align}
        &f(\lambda^{'}-\delta \lambda)=\sum_{i=1}^{3} f_i(\lambda^{'}-\delta \lambda)\\
        &f_i(\lambda^{'}-\delta \lambda)= \frac{a_i}{\sigma_i\sqrt{\pi}}\exp\left[-\left(\frac{\lambda^{'}-\delta \lambda-\lambda_{0}-\delta \lambda_i^{'}}{\sigma_i}\right)^2\right].
    \end{align}
    \label{eq:app1}
\end{subequations}
The values of the parameters $a_i$, $\sigma_i$, and $\delta \lambda_i^{'}$ can be found in \citet{Auchere2005a}. Then, the Doppler dimming term can be replaced by
\begin{equation}
    \begin{aligned}
    F_D=\int \sum_{i=1}^{3} f_i(\lambda^{'}-\delta \lambda)~\Phi(\lambda^{'}- \lambda_0, \boldsymbol{n})~d\lambda^{'}
    \end{aligned}
    \label{eq:app2}
\end{equation}
According to \citet{Bromiley2014}, the product between two normalized Gaussian functions is a normalized Gaussian function, which can be expressed as 
\begin{equation}
    \begin{aligned}
    &f(\lambda^{'},\mu_1,\sigma_1)~\Phi(\lambda^{'},\mu_2,\sigma_2)=\frac{S_{12}}{\sqrt{\pi}\sigma_{12}}\exp \left[-\left(\frac{\lambda^{'}-\mu_{12}}{\sigma_{12}}\right)^2\right],
    \end{aligned}
    \label{eq:app3}
\end{equation}
where $\sigma_{12}^2=\frac{\sigma_1^2\sigma_2^2}{\sigma_1^2+\sigma_2^2}$, $\mu_{12}=\frac{\mu_1\sigma_2^2+\mu_2\sigma_1^2}{\sigma_1^2+\sigma_2^2}$, and $S_{12}=\frac{1}{\sqrt{\pi(\sigma_1^2+\sigma_2^2)}}\exp \left[-\frac{(\mu_1-\mu_2)^2}{\sigma_1^2+\sigma_2^2} \right]$. For Gaussian profiles $f_i(\lambda^{'}-\delta \lambda)$ and $\Phi(\lambda^{'}- \lambda_0, \boldsymbol{n})$, $\mu_1=\lambda_0+\delta\lambda+\delta\lambda_i^{'}, \mu_2=\lambda_0, \sigma_1=\sigma_{i}, \sigma_2=\sqrt{2}\sigma_{\lambda}$. Then, the integration of Equation \ref{eq:app2} is 
\begin{equation}
    \begin{aligned}
    F_D=\sum_{i=1}^{3} \frac{a_i}{\sqrt{\pi}\sqrt{2\sigma_{\lambda}^2+\sigma_i^2}} \exp\left[ -\frac{(\delta \lambda_i^{'}+\frac{\lambda_0}{c}\boldsymbol{u} \cdot \boldsymbol{n})^2}{2\sigma_{\lambda}^2+\sigma_i^2}\right] 
    \end{aligned}
    \label{eq:app4}
\end{equation}
Finally, Equation \ref{eq:a0} can be expressed by
\begin{equation}
    \begin{aligned}
        j_r &=\frac{0.833h~B_{12}~\lambda_{0}}{4~\pi}~n_e~R_{HI}(T_e) \\
            &\times\int_{\phi}\int_{\varphi}\frac{1}{4\pi}\frac{11+3cos^2{\theta}}{12}~I_{disk} (\phi,\varphi)~R_{\odot}^2~(cos\gamma/d^2)~cos\phi \\ 
            &\times \sum_{i=1}^{3} \left\{\frac{a_i}{\sqrt{\pi}\sqrt{2\sigma_{\lambda}^2+\sigma_i^2}} \exp\left[ -\frac{(\delta \lambda_i^{'}+{\lambda_0~u\cos\beta}/{c})^2}{2\sigma_{\lambda}^2+\sigma_i^2}\right]  \right\}~d\phi~d\varphi.
    \end{aligned}
\end{equation}
From the data cube of the MHD simulation, we know all parameters, including plasma positions, angles, speeds, temperatures, and electron number densities. Thus, we can computer the emissivity of each plasma, and then obtain the 2D synthetic \hi \lya image by integrating along the $x$-axis, according to Equation \ref{eq:I1}.


\begin{ethics}
\begin{conflict}
The authors declare that they have no conflicts of interest.
\end{conflict}
\end{ethics}


\bibliographystyle{spr-mp-sola}
\bibliography{sola_bibliography_example}  

\IfFileExists{\jobname.bbl}{} {\typeout{}
\typeout{****************************************************}
\typeout{****************************************************}
\typeout{** Please run "bibtex \jobname" to obtain} \typeout{**
the bibliography and then re-run LaTeX} \typeout{** twice to fix
the references !}
\typeout{****************************************************}
\typeout{****************************************************}
\typeout{}}

\end{document}